\documentclass[a4paper,10pt,twocolumn]{article}
\usepackage{flushend}
\usepackage[left=1.8cm, right=1.5cm, top=2cm, bottom=2.5cm]{geometry}
\usepackage[T1]{fontenc}
\usepackage{lmodern}
\usepackage{amssymb,amsmath}
\usepackage[latin1]{inputenc}
\usepackage[english]{babel}
\usepackage{graphicx}
\usepackage[super, comma, sort&compress]{natbib}
\usepackage{hyperref}
\hypersetup{
  pdfauthor={Emmanuel Guilmeau, Maria Poienar, Stefan Kremer, Sylvain Marinel, Sylvie H\'ebert, Raymond Fr\'esard, Antoine Maignan},
  pdftitle={Mg substitution in CuCrO2 delafossite compounds},
  pdfsubject={doi:10.1016/j.ssc.2011.08.023},
  pdfkeywords={thermoelectricity, microstructure, low-dimensional oxides},
  pdfproducer={Copyright \textcopyright ~(2011) Emmanuel Guilmeau, Maria Poienar, Stefan Kremer, Sylvain Marinel, Sylvie H\'ebert, Raymond Fr\'esard, and Antoine Maignan. All rights reserved.},
  pdfborder={0 0 0}
}

\renewcommand{\thesection}{\Roman{section}}
\makeatletter
\renewcommand{\section}{
	\@ifstar\sectionStar\sectionNoStar
}
\makeatother
\newcommand{\sectionStar}[2]{
 	\par
	\vspace{1em}
	\pagebreak[3]
	\begin{center}
		\textbf{\MakeUppercase{#1}}
		\addcontentsline{toc}{section}{#1}
	\end{center}
	\vspace{1em}
}
\newcommand{\sectionNoStar}[1]{
	\par
	\vspace{1em}
	\pagebreak[3]
	\refstepcounter{section}
	\begin{center}
		\textbf{\thesection .~\MakeUppercase{#1}}
		\addcontentsline{toc}{section}{#1}
	\end{center}
	\vspace{1em}
}
\renewcommand{\thesubsection}{\Roman{section}.\Alph{subsection}}
\renewcommand{\subsection}[1]{
	\par
	\vspace{5pt}
	\refstepcounter{subsection}
	\begin{center}
		\normalsize\textbf{\thesubsection .~{#1}}
		\addcontentsline{toc}{subsection}{#1}
	\end{center}
	\vspace{5pt}
}


\begin{document}

	\title{~\\[-3em]
		\begin{picture}(0,0)(0,0)
			\setlength{\unitlength}{\textheight}
			\put(0,-1){\makebox(0,0)[c]{\tiny Copyright \textcopyright ~(2011) Emmanuel Guilmeau, Maria Poienar, Stefan Kremer, Sylvain Marinel, Sylvie H\'ebert, Raymond Fr\'esard, and Antoine Maignan. All rights reserved.}}
		\end{picture}~\\[-0.9em]
		\large \textbf{Mg substitution in CuCrO$_{\mathbf{2}}$ delafossite compounds}
	}
	\author{Emmanuel Guilmeau\textsuperscript{1},
		Maria Poienar\textsuperscript{1},
		Stefan Kremer\textsuperscript{1,2},
		Sylvain Marinel\textsuperscript{1},
		Sylvie H\'ebert\textsuperscript{1},\\
		Raymond Fr\'esard\textsuperscript{1,\!\!}
			\footnote{Corresponding author\quad
			E-mail: Raymond.Fresard@ensicaen.fr}
		~and Antoine Maignan\textsuperscript{1}\\[0.5em]
		\textit{\normalsize\textsuperscript{1}Laboratoire CRISMAT, UMR CNRS-ENSICAEN 6508, 14050 Caen, France}\\
		\textit{\normalsize\textsuperscript{2}Institut f\"ur Theorie der Kondensierten Materie, Karlsruhe Institute of Technology (KIT),}\\
		\textit{\normalsize 76128 Karlsruhe, Germany}\\[0.5em]
		{\normalsize (doi:10.1016/j.ssc.2011.08.023)}\\
	}

	\date{\normalsize
	\parbox{0.85\textwidth}{
		A detailed investigation of the series CuCr$_{1-x}$Mg$_{x}$O$_2$ ($x =0.0-0.05$) has been performed by making high-temperature resistivity and thermopower measurements, and by performing a theoretical
		analysis of the latter.
		Microstructure characterization has been carried out as well.
		Upon Mg$^{2+}$ for Cr$^{3+}$ substitution, a concomitant decrease in the electrical resistivity and thermopower values is found, up to $x \sim 0.02-0.03$, indicating a low solubility limit of Mg in the structure.
		This result is corroborated by scanning electron microscopy observations, showing the presence of MgCr$_2$O$_4$ spinels as soon as $x = 0.005$.
		The thermopower is discussed in the temperature-independent correlation functions ratio approximation as based on the Kubo formalism, and the dependence of the effective charge carrier density on the nominal Mg substitution rate is addressed.
		This leads to a solubility limit of 1.1\% Mg in the delafossite, confirmed by energy dispersive X-ray spectroscopy analysis.
	}}

	\maketitle\thispagestyle{empty}
	\newpage
	\pagestyle{headings}

\section{Introduction} \label{sec:intro}

	In recent years, there has been a renewed interest in delafossite compounds of general formula CuMO$_2$, mainly concerning either fundamental research such as frustrated magnetic interactions in triangular lattices \cite{Kimura2006,Kadowaki90,Olariu2006,Ye2006,Collins97,mendels2004,Li2011}, or the search for new materials such as transparent p-type conducting oxides \cite{kawazoe1997, ingram2004, Nagarajan2001,Snure2007,Scanlon2009} and more
	recently thermoelectric materials \cite{koumoto2001, hayashi2007, okuda2005, park2007, ono2007, Ru2011}.
	In this context, understanding the relationships between chemical composition and structural and thermoelectric properties is very interesting, and it can be helpful in order to design and optimize delafossite compounds for their integration in potential applications.
	Recently, several groups have reported interesting results on the thermoelectric properties of CuCrO$_2$ bulk compounds.
	A very striking result follows from low Mg$^{2+}$ doping:
	it strongly impacts the transport properties, without inducing changes in the magnetic structure as shown by a neutron diffraction study \cite{Poienar2009}.
	However, it must be recalled that Mg substitution in CuCrO$_2$ not only results in the formation of a delafossite CuCr$_{1-x}$Mg$_{x}$O$_2$ phase, but also in the formation of an impurity MgCr$_{2}$O$_4$ spinel phase, and eventually of a ternary CuO phase (for large enough substitution rate), as attested by an X-ray diffraction (XRD) study \cite{Maignan2009}.
	Therefore, even for low Mg substitution rate, the Mg doping level of the delafossite differs from the Mg nominal content.
	In addition, this makes it difficult to determine the solubility limit of Mg in CuCrO$_2$, and these points deserve further study.
	In order to shed light on this problem and to better characterize the effect of Mg substitution in CuCrO$_2$, in this paper we first report the high temperature thermoelectric properties, i.\ e., the thermopower and the electrical resistivity.
	We have measured between room temperature and 1100\ K, namely in a temperature range in which thermoelectric oxides have proved useful.
	Second, the thermopower data are analyzed in the framework of the temperature-independent correlation functions ratio (TICR) approximation \cite{RF02}, using an effective density of states \cite{KF11}, allowing us to address the dependence of the effective charge carrier density on the nominal Mg doping.
	Third, this provides us with an estimate of the solubility limit of Mg in CuCrO$_2$.
	Finally, we further confirm this estimate by energy dispersive X-ray spectroscopy (EDS) analysis.

\section{Sample preparation and characterization} \label{sec:prep}

	All samples, belonging to the CuCr$_{1-x}$Mg$_{x}$O$_2$ series, were prepared using a standard solid-state reaction route.
	The starting powders, Cu$_2$O (Aldrich, 99\%), Cr$_2$O$_3$ (Alfa Aesar, 99\%), and MgO (Cerac, 99.95\%), were weighed in stoichiometric amounts and ground together by ball milling.
	The resulting powders were pressed uniaxially under $300\,\mathrm{MPa}$, using polyvinyl alcohol binder to form parallelepipedic $3 \times 3 \times 12 $\,mm$^3$ samples.
	These were then sintered at $1200^\circ$C for 24h in air on platinum foil to avoid any contamination from the alumina crucible.
	The scanning electron microscopy (SEM) observations were made using a FEG Zeiss Supra 55.
	The cationic compositions were determined by energy dispersive X-ray spectroscopy (EDX) EDAX.
	The electrical conductivity and thermopower were measured simultaneously using an ULVAC-ZEM3 device between 50 and $800^\circ$C under helium.

\section{Results and discussion} \label{sec:Results}

\begin{figure}[t]
		\centering
		\includegraphics[width=\columnwidth,clip]{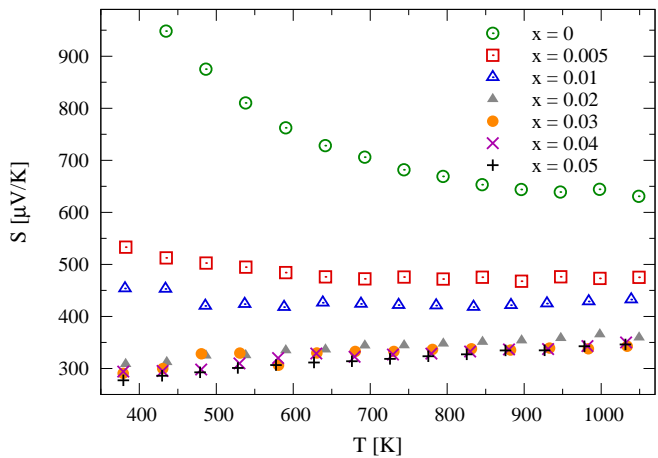}
		\caption{
			Temperature dependence of the thermopower in the series CuCr$_{1-x}$Mg$_{x}$O$_2$.
		}
		\label{fig:tep}
	\end{figure}

	As the thermopower ($S$) of CuCr$_{1-x}$Mg$_{x}$O$_2$ strongly increases with temperature below but in the vicinity of room temperature for $x\geq 1\%$ \cite{Maignan2009}, and hence shows a promising potential for high-temperature applications, we extended the transport measurements up to 1100\ K.
	Also, as we show below, such thermopower measurements are useful for determining the solubility limit of Mg in the delafossite CuCr$_{1-x}$Mg$_{x}$O$_2$.
	As seen in Fig.~\ref{fig:tep},  positive $S$ values for all $x$ are found with a magnitude that decreases as $x$ increases up to $x=0.03$, where it saturates.
	This indicates an increase in the hole concentration in the nominal CuCr$_{1-x}$Mg$_{x}$O$_2$ series before saturation.
	Ono et al. \cite{ono2007} interpreted this variation in terms of hole doping at the Cr sites by the partial substitution of Mg$^{2+}$ for Cr$^{3+}$.
	However, as the insulating spinel phase appears to form at $x=0.01$ based on XRD measurements \cite{Maignan2009}, the decrease in Seebeck coefficient and electrical resistivity cannot be simply attributed to the substitution of Cr$^{3+}$ by Mg$^{2+}$ in the delafossite phase, but at most to a partial substitution, which we address below.

\begin{figure}[t]
		\centering
		\includegraphics[width=\columnwidth,clip]{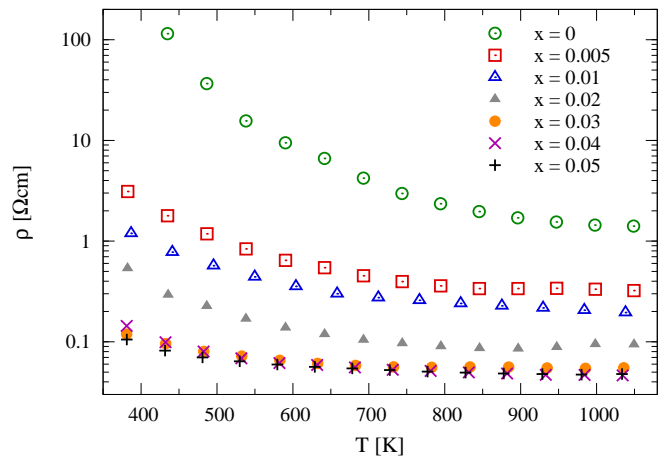}
		\caption{
			Temperature dependence of the electrical resistivity in the series CuCr$_{1-x}$Mg$_{x}$O$_2$.
		}
		\label{fig:rho}
	\end{figure}

	Figure~~\ref{fig:rho} shows the temperature dependence of the electrical resistivity, $\rho$ of CuCr$_{1-x}$Mg$_{x}$O$_2$ samples ($0 \leq x \leq 0.05$).
	As previously reported \cite{okuda2005,ono2007,Maignan2009}, the $\rho(T)$ curves of the CuCr$_{1-x}$Mg$_{x}$O$_2$ series exhibit a typical semiconducting behavior over the whole temperature range.
	With increasing doping level up to $x = 0.03$, $\rho$ decreases monotonically, whereas, above $x=0.03$, the $\rho(T)$ curves are clearly $x$-independent.
	The change of the thermopower with temperature and doping level follows the same behavior, with a minimum reached for $x\geq 0.03$.
	The $S(T)$ curve of CuCrO$_2$ ($x = 0$) has a negative $\mathrm{d}S/\mathrm{d}T$ value; however, with Mg doping, a change in the temperature dependence is observable with $\mathrm{d}S/\mathrm{d}T >0$  for the higher Mg contents.
	Such temperature and doping dependences of $\rho$ and $S$ are in very good agreement with the previous study by Ono et al. \cite{ono2007}.
	The Seebeck coefficient results from the charge carrier concentration (Cr$^{4+}$ per available chromium ion site) may be analyzed by means of a generalized Heikes formula \cite{koshibae}.
	However, we obtained that the Heikes formula including a spin and orbital degeneracy term yields dissatisfying results.
	In fact, when compared with measured Seebeck values, they are systematically larger.
	Ono et al. mentioned in their study a good agreement between the Heikes model and their variation of the Seebeck coefficient versus the nominal Mg fraction at $1100\,\mathrm{K}$.
	However, they considered the spin entropy terms $g_3 = 4$ and $g_4 = 9$, which yield $70\,\mathrm{\mu V/K}$, although experimentally $S$ saturates at about $310\,\mathrm{\mu V/K}$ for $x >0.03$ (see Fig.~\ref{fig:tep}).
	As the spinel phase is formed in their compounds with $x \geq 0.03$, and taking into account from the present study that $x$ cannot exceed 0.015, the values of Seebeck coefficient reported in their paper for $x = 0.03$, 0.04 and 0.05 are obviously not correct, or should be attributed to smaller $x$ fractions.
	This therefore calls for a more sophisticated explanation.

\section{Theory} \label{sec:theory}

	In order to shed more light on the above results, we now analyze the thermopower data in the temperature-independent correlation function ratio (TICR) approximation \cite{RF02}.
	In this framework, the thermopower is represented as
	\begin{equation}
		S(T) = \frac{1}{q_eT} \big( E_0 - \mu(T) \big) \;,\label{eq:ticr}
	\end{equation}
	with $q_e$ the (negative) electron charge, and $E_0$ the ratio of the two correlation functions entering the expression of the thermopower in Kubo formalism \cite{Luttinger,Mahan}.
	Originally devoted to electron-doped manganites \cite{RF02,Chemat}, this approach has recently been extended to hole-doped delafossites \cite{Maignan09b,KF11}, in which case it gave an excellent account of the thermopower of CuRhO$_2$.
	In the course of the analysis an apparent Fermi liquid (AFL) behavior has been identified, which is characterized by a thermopower which behaves as
	\begin{equation}
		S(T) \simeq S_0 + \frac{\varpi}{|q_e|} T \,.\label{eq:safl}
	\end{equation}
	The temperature range where the AFL is observed, the coefficients $S_0$ and $\varpi$ are doping dependent.%,
	In contrast, in a Fermi liquid, $S_0$ vanishes.
	The precise value of these coefficients can be obtained from an effective density of states (DOS).
	This may take the form
	\begin{equation}
		\rho(\epsilon) = (\rho_0 + \epsilon \rho^{\prime})\Theta(-\epsilon)\,,\label{eq:dos}
	\end{equation}
	which accounts for the discontinuous behavior of the DOS at the upper band edge evidenced in first-principles calculations \cite{Eyert081,Eyert082,Singh07,Maignan09b}.
	Since the materials we consider are hole doped, the slope $\rho^{\prime}$ is negative.

	\begin{table}[t!]
		\centering
		\begin{tabular}{@{}l@{\,}|@{\,}c@{\,}|@{\,}c@{\,}|@{\,}c@{\,}|@{\,}c@{\,}|@{\,}c@{\,}|@{\,}c@{}}
			\hline
			$x$& $\rho_0$& $\rho^{\prime}$        & $E_0$ & $\mathrm{\bar x}$ & ${\bar \rho}_0$ & ${\bar \rho}\,^{\prime}$\\
			~     &[(eV)${}^{-1}$]&[(keV)${}^{-2}$]& [meV] &                   & [(keV)${}^{-1}$]& [(eV)${}^{-2}$]\\ \hline
			0.5\% &      175  &     35 &     131   & 0.2\% &  70 & 14\\
			1.0\% &      141  &     28 &     153   & 0.6\% &  85 & 17\\
			2.0\% &      220  &     44 &     112   & 1.1\% & 121 & 24\\
			3.0\% &      608  &    122 &      58.2 & 1.2\% & 243 & 49\\
			4.0\% &      823  &    165 &      51.9 & 1.45\% & 298 & 60\\
			5.0\% &     1180  &    236 &      36.7 & 1.14\% & 330 & 66\\ \hline
		\end{tabular}
		\caption{
			Parameters of the effective density of states as determined by a fit of Eq.~(\ref{eq:boltzmann}) to the experimental values of CuCr$_{1-x}$Mg$_{x}$O$_2$ given for the values of the nominal doping $x$ as well as for those obtained from EDS measurement ($\mathrm{\bar x}$, ${\bar \rho_0}$, ${\bar\rho}\,^{\prime}$) (see Fig.~\ref{fig:mgcont} below).
		}
		\label{tab:expvstheo}
	\end{table}
	From this DOS, the doping $x$ can be calculated for a given chemical potential $\mu$ as
	\begin{equation}
		x = \int_{-\infty}^{\infty}\mathrm{d}\epsilon\big(1 - f(\epsilon -\mu)\big)\rho(\epsilon) \label{eq:xdemu}
	\end{equation}
	where $f(x)$ represents the Fermi function.
	Therefore, combining the chemical potential extracted by the TICR framework Eq.~(\ref{eq:ticr}) and a common effective DOS for all samples of an experimental series, the exact doping can be determined.
	This means that this approach can be viewed as an alternative way to identify the effective charge carrier density, which is complementary to the EDS measurements discussed below.

	In order to extract the DOS from the temperature dependence of the chemical potential of one sample, Eq.~(\ref{eq:xdemu}) has to be inverted.
	In the Boltzmann approximation, the chemical potential $\mu_B$ is given by
	\begin{equation}
		\mu_{Bz} = k_{\mathrm{B}}T\ln\!\left(\frac{k_{\mathrm{B}}T (\rho_0  - \rho^{\prime} k_{\mathrm{B}}T)}{x}\right)\label{eq:boltzmann}.
	\end{equation}
	This formula has proven to show a stable quadratic expansion at high temperatures \cite{KF11}, describing an AFL behavior.
	However, since the TICR constant $E_0$ usually differs from the Fermi energy it produces an additional hyperbolic offset which is negligible only at very high temperatures.
	For lower temperatures, it can still be taken into account by considering the difference of the chemical potential with respect to the relative temperature $T-T_1$.
	This means that the quantity
	\begin{align}
		\frac{TS(T)-T_1S(T_1)}{T-T_1} &= \frac{\mu(T) - \mu(T_1)}{T-T_1}\nonumber\\
		&= \frac{\varpi T_1}{|q_e|}-S_0 + \frac{\varpi}{|q_e|}T
	\end{align}
	is independent of the hyperbolic offset, and therefore shows a linear behavior.

	\begin{figure}[t!]
		\centering
		\includegraphics[width=\columnwidth]{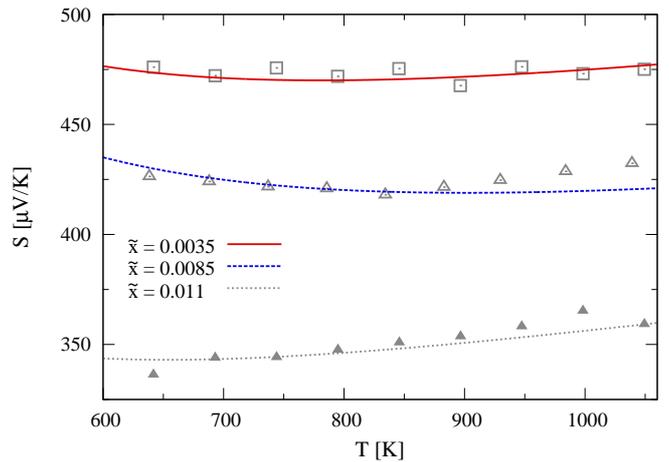}
		\caption{
			Comparison of the thermopower given by experimental data (points) for nominal doping values ${x} = 0.5\%$ (solid), $1\%$ (dashed), and $2\%$ (dotted), to that obtained from the parameters of the density of states for values of the doping of ${\tilde {x}} = 0.0035$ (solid), $0.0085$ (dashed), and $0.011$ (dotted).
		}
		\label{fig:Stheo}
	\end{figure}
	Carrying out the analysis for CuCr$_{1-x}$Mg$_{x}$O$_2$ (Tab.~\ref{tab:expvstheo}) results in TICR parameters $E_0$ close to those reported for doped manganites \cite{RF02} and in slopes $\rho^{\prime}$ dominating the effective DOS, which is in agreement with a previous first-principles study \cite{Maignan2009}.
	However, having the effective DOS dominated by the slope challenges the determination of the small discontinuity $\rho_0$.
	Furthermore, the results exhibit a strong increase of both parameters for the three highest doped samples.
	Since this effect might follow from the formation of the secondary spinel or CuO phase as previously discussed, these samples are omitted in the following discussion.

	The variations in the remaining parameters suggest that the effective charge carrier density differs from that for the nominal doping.
	Indeed, the parameters obtained were found to be in better agreement.
	However, deviations still occur for the lowest doped samples.
	In order to determine the effective charge carrier density from a common DOS, the parameters of the sample with nominal doping $x=2\%$
	\begin{equation}
		\rho_0 = 0.121\,\mathrm{(eV)^{-1}}\qquad \rho^{\prime} = -24.2\,\mathrm{(eV)^{-2}}
	\end{equation}
	lead to doping values of
	\begin{equation}
		{\tilde {x}} = 0.35\%,\ 0.85\%, 1.1\%
	\end{equation}
	for ${x} = 0.5\%$, $1\%$, and $2\%$, respectively.
	Using these values and performing the exact numerical solution of Eq.~(\ref{eq:xdemu}) in order to calculate the thermopower leads to good agreement between theory and experiment as can be seen in Fig.~\ref{fig:Stheo}.
	This suggests that the solubility limit of Mg in the delafossite CuCr$_{1-x}$Mg$_{x}$O$_2$ is close to $1.1\%$.

\section{Microscopy} \label{sec:microscopy}

\begin{figure}[t]
		\centering
		\includegraphics[width=0.7\columnwidth,clip]{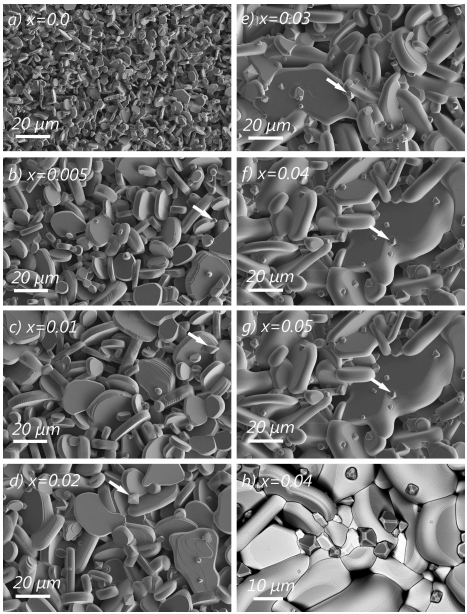}
		\vspace{1em}
		\caption{
			SEM micrographs of CuCr$_{1-x}$Mg$_{x}$O$_2$ ($x = 0$, 0.005, 0.01, 0.02, 0.03, 0.04, 0.05) samples.
			Platelike grains are typical of the layered delafossite structure.
			Small grains of octahedral shape correspond to the spinel phase MgCr$_2$O$_4$ (indicated with arrows).
			CuO and spinel phases are shown respectively as white and dark contrasts in figure $h$ ($x = 0.04$).
		}
		\label{fig:sem}
		\vspace{2em}
	\end{figure}

	In order to further clarify the effect of the Mg substitution, all the samples in the CuCr$_{1-x}$Mg$_{x}$O$_2$ series were observed by SEM, and statistical EDS analyses on a minimum of 20 grains of the different compounds were carried out to determine the cationic compositions of each grain.
	Fig.~\ref{fig:sem} shows the micrographs of the CuCr$_{1-x}$Mg$_{x}$O$_2$ compounds at different Mg substitution levels.
	As $x$ increases up to 0.04, the size of the plate-like grains, characteristic of the layered delafossite structure, systematically increases.
	The average diameter for the CuCrO$_2$ phase increases from $\sim 5\,\mathrm{\mu m}$ up to $\sim 100\,\mathrm{\mu m}$ in the highly Mg-doped compounds, indicating a change in the cationic or anionic diffusion.
	The spinel phase, identified as smaller grains of octahedral shape, appears in the Mg-substituted samples and the amount increases with $x$ (for $0.005 \leq x \leq 0.05$).
	The presence of CuO phase observed in the diffraction patterns for $x \geq 0.04$ \cite{Maignan2009} is evidenced by the small white contrast grains in the CuCr$_{1.96}$Mg$_{0.04}$O$_2$ compound (Fig.~\ref{fig:sem}h).

	\begin{figure}[t]
		\centering
		\includegraphics[width=0.9\columnwidth]{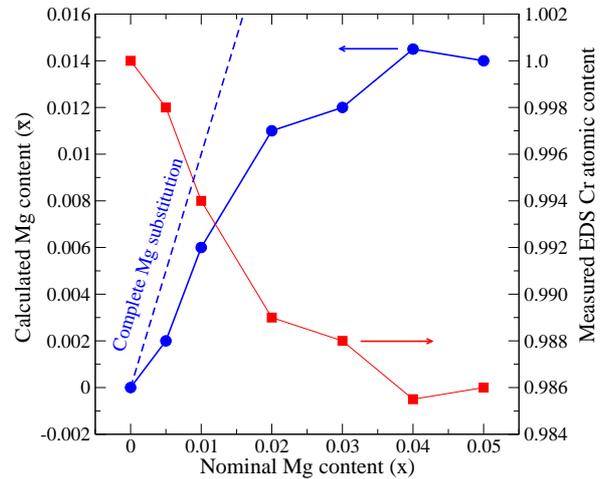}
		\caption{
			Calculated Mg content (left Y-axis) and measured EDS Cr cationic content (right Y-axis) versus nominal Mg content ($x$) in the CuCr$_{1-x}$Mg$_{x}$O$_2$ series.
		}
		\label{fig:mgcont}
	\end{figure}
	The black octahedral grains correspond to the spinel phase.
	EDS analyses confirmed that the spinel phase contains Mg and Cr, but also Cu in a smaller amount.
	Normalizing the Mg content to unity, the atomic ratios of Mg, Cr and Cu are 1, 1.82, and 0.31, respectively.
	Moreover, EDS analyses performed on each individual plate-like grain confirmed the presence of Mg in the delafossite structure as soon as $x=0.005$.
	However, due to the detection limit and the resolution of the apparatus, the Mg cationic content could not be determined quantitatively.
	Based on a qualitative approach, the analyses performed on delafossite grains reveal that the energy peak intensity of the MgK$_\alpha$ line increases as the doping level increases and reaches the maximum for $x = 0.03$.
	This result evidences the successful substitution of Cr by Mg, which is lower than the nominal value, as a Mg-based spinel is formed as soon as $x = 0.005$.
	To confirm this qualitative information, the chromium and copper cationic compositions have been determined for each composition in a panel of 20 grains.
	Fig.~\ref{fig:mgcont} shows the evolution of the measured EDS Cr atomic content (right Y-axis) versus the nominal Mg substitution rate $x$.
	Normalizing the Cu content to unity, the Mg cationic content ($\mathrm{\bar{x}}$) has been calculated and is drawn in the same figure (left Y-axis).
	The actual Cr content decreases as the nominal Mg substitution rate increases, which suggests substitution of Cr by Mg in the delafossite structure.
	As shown clearly from the deviation from the nominal value, the substitution is not complete due to the formation of the spinel phase.
	It must be emphasized that the Mg solubility determined here is much lower than previously reported \cite{okuda2005,ono2007}, but it is in good agreement with the above theoretical analysis.
	This very low amount of substitution also explains the limited variation of the unit cell parameters with the substitution \cite{Maignan2009}.
	Remarkably, this small amount is sufficient to strongly affect the transport properties.
	However, one cannot exclude that this substitution is also accompanied by other hardly measurable non-stoechiometry phenomena such as tiny changes in the Cu/Cr ratio and/or the oxygen
	content.

\section{Conclusion}

	In summary, Mg-doped CuCrO${}_2$ has been investigated by SEM, EDS, electrical conductivity, and thermopower measurements.
	Each method showed abnormal behavior for samples with nominal doping $x=0.03$ and above, which was indicated by a theoretical study based on the data of the thermopower, too.
	For values of the doping above this, a CuO phase is observed, therefore limiting the Mg solubility in the delafossite structure.
	Furthermore, the renormalized doping values from EDS as well from the theoretical study are found to be much lower than the nominal ones.
	This was represented in the SEM observations by the forming of the secondary spinel phase, in contrast to previously published work \cite{ono2007}.
	However, the obtained values of the doping clearly indicate the substitution of Cr by Mg resulting in peculiar hole doping of the rather narrow Cr 3d bands.
	In spite of the small change of their charge carrier density, the transport properties show a very sensitive dependence.
	This behavior, in particular that of the thermopower, was shown to be very well described within the TICR framework applied to an effective density of states and treated in the Boltzmann regime.

\section*{Acknowledgments}

	We are grateful to Christine Martin for enlightening discussions.
	This work was supported by the ANR through ANR-08-BLAN-0005-01, by the R\'egion Basse-Normandie, and by the Minist\`ere de la Recherche.
	SK acknowledges financial support by the Research Unit 960 ``Quantum Phase Transitions'', and SK and RF by the FGU.

\end{document}